\pdfoutput=1
\documentclass{article}

\usepackage[final]{neurips_2022_ml4ps}

\usepackage[utf8]{inputenc} 
\usepackage[T1]{fontenc}    
\usepackage{hyperref}       
\usepackage{url}            
\usepackage{booktabs}       
\usepackage{amsfonts}       
\usepackage{nicefrac}       
\usepackage{microtype}      
\usepackage{xcolor}         
\usepackage{gensymb}
\usepackage{soul}
\usepackage{graphicx}
\usepackage{amsmath}
\usepackage{orcidlink}

\title{SuNeRF: Validation of a 3D Global Reconstruction of the Solar Corona Using Simulated EUV Images}

\author{
  Kyriaki-Margarita Bintsi\(^*\) \orcidlink{0000-0003-3875-7757} \\
  Imperial College London\\
  London SW7 2AZ, UK \\
  \texttt{m.bintsi19@imperial.ac.uk} \\
  \And
  Robert Jarolim\(^*\) \orcidlink{0000-0002-9309-2981} \\
  University of Graz \\
  Universit\"atspl. 3, 8010 Graz, Austria \\
  \texttt{robert.jarolim@uni-graz.at} \\
  \AND
  Benoit Tremblay\(^*\) \orcidlink{0000-0002-5181-7913} \\
  High Altitude Observatory \\
  3080 Center Green Dr. \\ Boulder, CO 80301, USA \\
  \texttt{btremblay@ucar.edu} \\
  \And
  Miraflor Santos\(^*\) \\
  Massachusetts Institute of Technology \\
  77 Massachusetts Ave \\ Cambridge, MA 02139, USA \\
  \texttt{miras@mit.edu} \\
  \And
  Anna Jungbluth\(^*\) \orcidlink{0000-0002-9888-6262} \\
  University of Oxford \\
  Oxford OX1 2JD, UK \\
  \texttt{anna.jungbluth@physics.ox.ac.uk} \\
  \And
  James Paul Mason \orcidlink{0000-0002-3783-5509} \\
  Applied Physics Lab. \\ Johns Hopkins University \\
  11100 Johns Hopkins Rd. \\ Laurel, MD 20723, USA \\
  \texttt{james.mason@jhuapl.edu} \\
  \And
  Sairam Sundaresan \orcidlink{0000-0002-6648-0591} \\
  Intel Labs \\
  2200 Mission College Blvd. \\ Santa Clara, CA 95054, USA \\
  \texttt{sairam.sundaresan@intel.com} \\
  \And
  Angelos Vourlidas \orcidlink{0000-0002-8164-5948} \\
  Applied Physics Lab. \\ Johns Hopkins University \\
  11100 Johns Hopkins Rd. \\ Laurel, MD 20723, USA \\
  \texttt{Angelos.Vourlidas@jhuapl.edu} \\
  \And
  Cooper Downs \\
  Predictive Science Inc. \\
  9990 Mesa Rim Rd. Suite 170 \\ San Diego, CA 92121 \\
  \texttt{cdowns@predsci.com} \\
  \And
  Ronald M. Caplan \\
  Predictive Science Inc. \\
  9990 Mesa Rim Rd. Suite 170 \\ San Diego, CA 92121 \\
  \texttt{caplanr@predsci.com} \\
  \AND
  Andr\'es Mu\~noz-Jaramillo \orcidlink{0000-0002-4716-0840} \\
  Southwest Research Institute \\
  1050 Walnut St., Suite 300 \\ Boulder, CO 80302, USA \\
  \texttt{amunozj@boulder.swri.edu} \\
}

\begin{document}

\maketitle

\vspace{-0.7cm}
\begin{center}
    $^*$ These authors contributed equally to this work.
\end{center}
\vspace{0.7cm}

\begin{abstract}
Extreme Ultraviolet (EUV) light emitted by the Sun impacts satellite operations and communications and affects the habitability of planets. 
Currently, EUV-observing instruments are constrained to viewing the Sun from its equator  (i.e., ecliptic), limiting our ability to forecast EUV emission for other viewpoints (e.g. solar poles), and to generalize our knowledge of the Sun-Earth system to other host stars. 
In this work, we adapt Neural Radiance Fields (NeRFs) to the physical properties of the Sun and demonstrate that non-ecliptic viewpoints could be reconstructed from observations limited to the solar ecliptic.
To validate our approach, we train on simulations of solar EUV emission that provide a ground truth for all viewpoints. 
Our model accurately reconstructs the simulated 3D structure of the Sun, achieving a peak signal-to-noise ratio of 43.3 dB and a mean absolute relative error of 0.3\% for non-ecliptic viewpoints.
Our method provides a consistent 3D reconstruction of the Sun from a limited number of viewpoints, thus highlighting the potential to create a virtual instrument for satellite observations of the Sun. Its extension to real observations will provide the missing link to compare the Sun to other stars and to improve space-weather forecasting.
\end{abstract}

\section{Introduction}

The Sun enables life on Earth, but is also hazardous for humans and satellites in space. Extreme Ultraviolet light (EUV) changes thermospheric density, affecting atmospheric drag on satellites in low Earth orbit. 
Current EUV imaging instruments are tied to the ecliptic plane, i.e. they orbit and image the Sun from its equator, and no satellites currently observe the solar poles directly. Over the last 15 years, at most three satellites observed the Sun at any given time. However, a complete image of the Sun is required to forecast EUV radiation to protect our assets, and to relate the Sun to other stars in the universe that are observed as distant points in the sky with unknown inclination angles.

Reconstructing the 3D geometry of the Sun is very challenging. 
The outermost layer of the solar atmosphere, the solar corona, consists of partially-ionized plasma (i.e., hot gas) that emits light of all wavelengths, including in EUV. This radiative environment, made of a optically thin plasma, is very different than Earth's in which solid and optically thick objects mainly absorb and reflect light. In optically thin environments, observed intensities can originate from multiple points, which prevents the precise mapping to a single spatial location. 

Due to the lack of global 3D coronal reconstructions, current state-of-the-art 3D solar maps combine simultaneous images viewed by different observers to create synchronic maps \citep[e.g., ][]{Caplan2016}. These maps are assembled by reprojecting different viewpoints on to a spherical surface, assuming that the observed intensity originates from a single plane and neglecting the three-dimensionality of the solar corona. This leads to projection effects that are especially prominent near the limb (i.e., edge) of the solar disk and reduces the scientific utility of off-limb imaging.

To maximize the science return of multiple viewpoints, we propose a novel approach that unifies and smoothly integrates data from multiple perspectives into a consistent 3D representation of the solar corona. We leverage Neural Radiance Fields \citep[NeRFs: ][]{Mildenhall2020}, 
which are (fully-connected) neural networks that can achieve state-of-the-art 3D scene representation and generate novel views from a limited number of input images. 
We adapt standard NeRFs to match the physical reality of the Sun, introducing radiative transfer and geometric sampling to create a solar NeRF (SuNeRF). 
In the absence of observations from outside of the ecliptic, we evaluate our SuNeRF on simulated images of the Sun in EUV that provide ground truth information for all viewpoints.

\section{Data}
\label{S-Data}

\begin{figure}[t]
    \centering
    \noindent\includegraphics[width=0.33\textwidth]{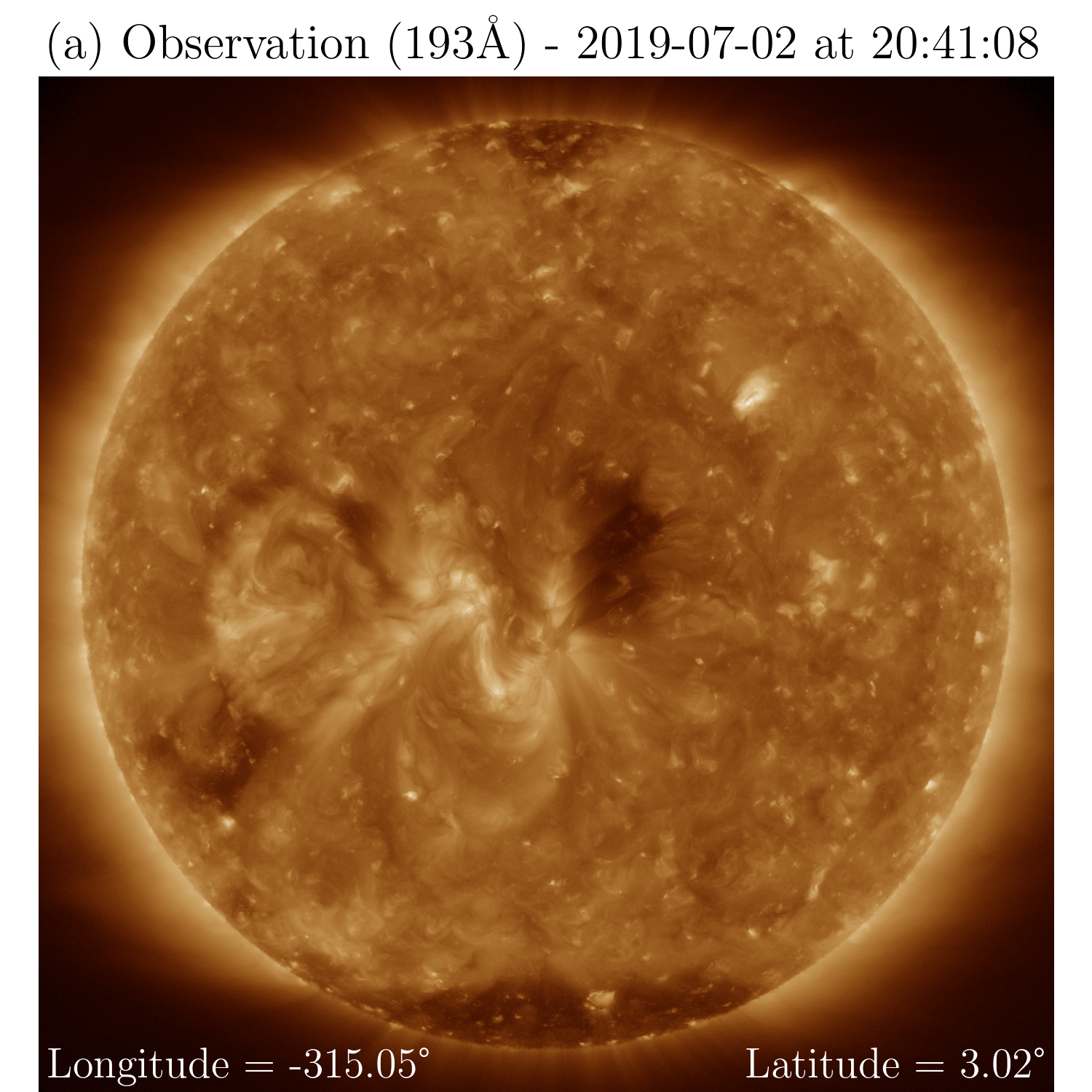}\includegraphics[width=0.33\textwidth]{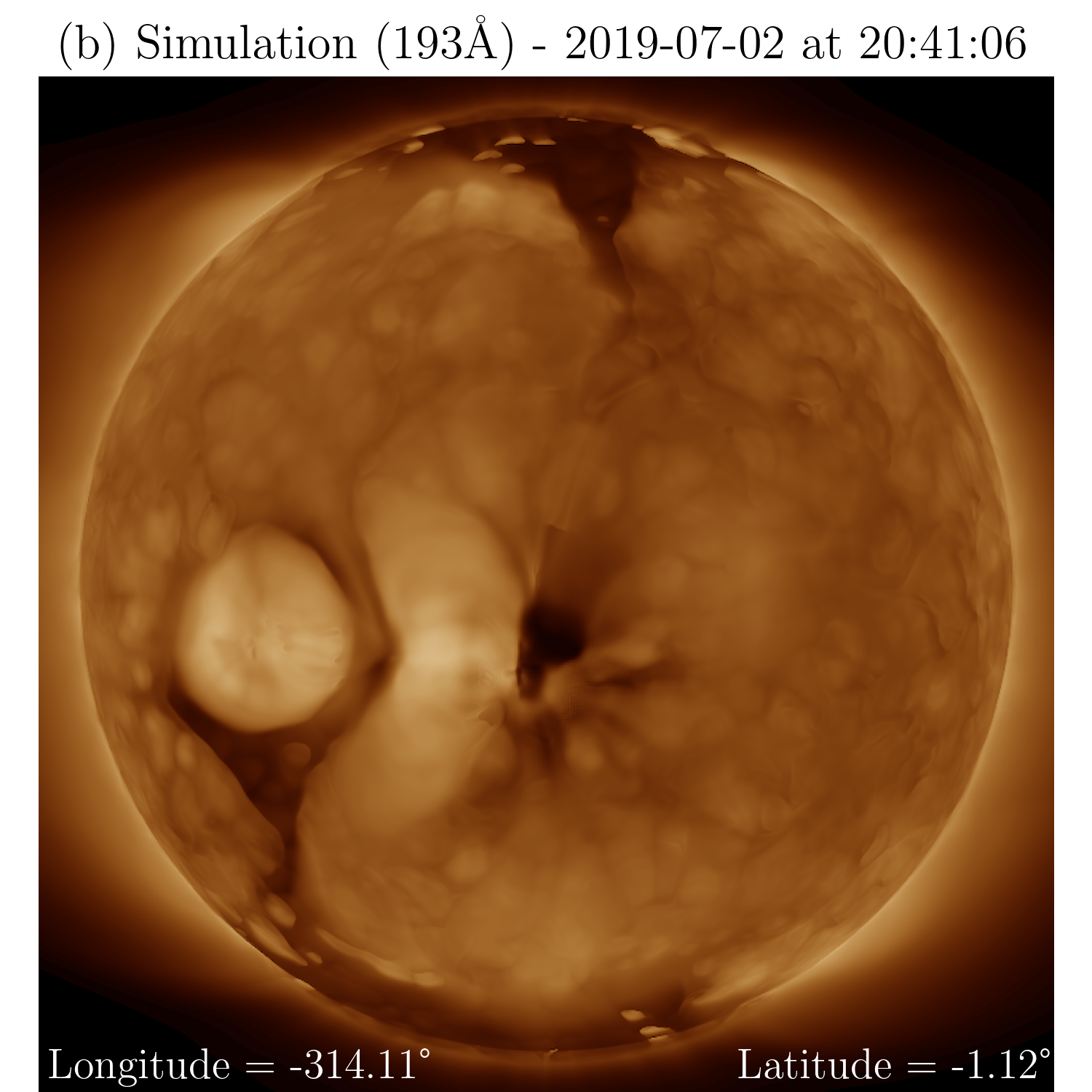}\includegraphics[width=0.33\textwidth]{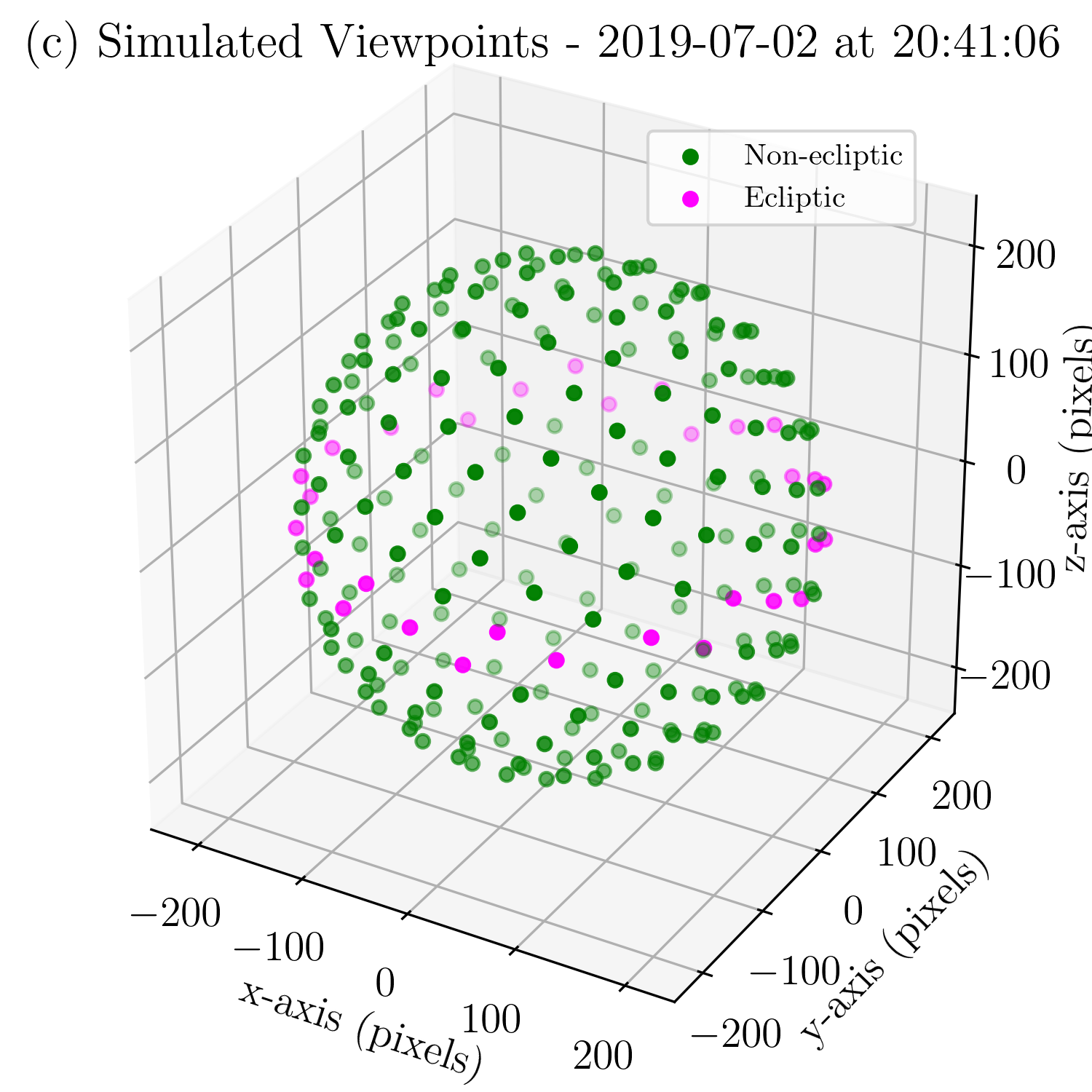}
    \caption{(a) Satellite image of the Sun at 193{\AA} that was captured from the ecliptic on 2019-07-02 at 20:41:08 (UT). (b) Simulated image of the satellite viewpoint (offsetted), extracted from a 3D model of the Sun. (c) Positions of the 256 viewpoints that were extracted from the 3D model. The color coding indicates which viewpoints were used for the training set (magenta) and the test set (green).} \label{dataset}
\end{figure}

In this work, we take advantage of a magnetohydrodynamic (MHD) simulation of the solar corona used to forecast the state of the solar atmosphere prior to the 2019-07-02 total solar eclipse by Predictive Science Inc.\ \citep[PSI; see the website\footnote{Predictive Science Inc. 2019-07-02 total solar eclipse prediction: \href{https://www.predsci.com/eclipse2019}{www.predsci.com/eclipse2019}.} and][for more details]{Boe2021, Boe2022}.  MHD simulations estimate the global 3D distribution of plasma parameters and magnetic field in the solar atmosphere, allowing one to forward model EUV images and other diagnostics from any viewpoint \citep{Mikic2018}. Figures \ref{dataset}(a) and \ref{dataset}(b) show the observed EUV image and data-constrained simulation respectively. The PSI model dataset consists of 256 forward modeled 193{\AA} images captured from evenly spaced viewpoints, providing full coverage of the simulated Sun (Figure \ref{dataset}(c)). We train our SuNeRF model on synthetic observations viewed from $|\text{latitudes}| \leq 7 \degree$ (i.e., 32 images viewed from the ecliptic), and we test the model on the the remaining viewpoints (i.e., images viewed from outside of the ecliptic). This idealized dataset allows us to test if the SuNeRF can successfully generate an accurate 3D reconstruction solely from images taken from the ecliptic.

\section{Methods: From NeRFs to SuNeRFs}

\begin{figure}[t]
    \centering
    \includegraphics[width = 0.8\textwidth]{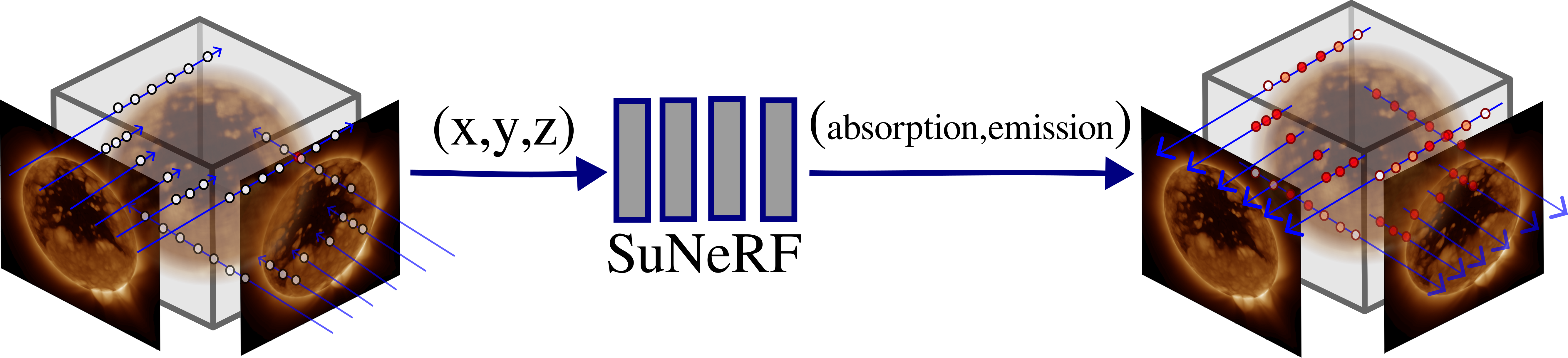}
    \caption{Overview of our SuNeRF model that is trained to reconstruct the geometry of the Sun from a set of training images. The simulation volume is represented by a neural network that maps each coordinate point $(x, y, z)$ to the corresponding emission and absorption coefficient ($\epsilon$, $\kappa$). For each pixel, we sample rays from the volume and compute total intensity using radiative transfer principles.}
    \label{pipeline}
\end{figure}

NeRFs \citep[][]{Mildenhall2020} learn the density and color at each point in a volume based on a set of images captured from different viewpoints. The resulting representation can then be used to render novel scenes. 
In this study, we apply two primary modifications to the NeRF model to account for the physical reality of the Sun and generate novel views of our star, which we refer to as a Sun NeRF (SuNeRF).  NeRFs were designed to represent solid or transparent objects. 
However, pixel values in images of the solar corona integrate emitting and absorbing (i.e., optically thin) plasma elements along the line-of-sight. 
Therefore, we replace the density and color predictions of NeRFs with emission and absorption. For each pixel, we construct the total emission by sampling points along the ray path. At each point $(x, y, z)$, our network predicts the emission and absorption coefficients ($\epsilon$, $\kappa$). The emission per point, $I$, is evaluated by multiplying the emission coefficient $\kappa$ with the spacing of the sampled ray $ds$. The absorption per point $A$ is, by definition, scaled between 0 and 1 (total and no absorption, resp.) using $A(x, y, z) = \exp{(-\kappa \, \, ds)}$.  The total observed intensity $\hat{I}$ is then computed by integrating all sampled points, where each intensity value is reduced by absorption along the path of propagation (from the origin to the observer): 
\begin{equation}
    \widehat{I} = \sum_k I_k * \prod_{i}^{k-1} A_i,
\end{equation}
where the indices refer to the sampled points along the ray.
For the final pixel value, we use an $asinh$ stretching to optimize the value range for training. We also adapt the NeRF ray sampling to account for the geometry of the Sun. We sample rays ranging from [-1.3, 1.3] solar radii from the Sun.

\begin{figure}[t]
    \centering
    \includegraphics[width = 0.94\textwidth, trim={0 0.22cm 0 0.01cm}, clip]{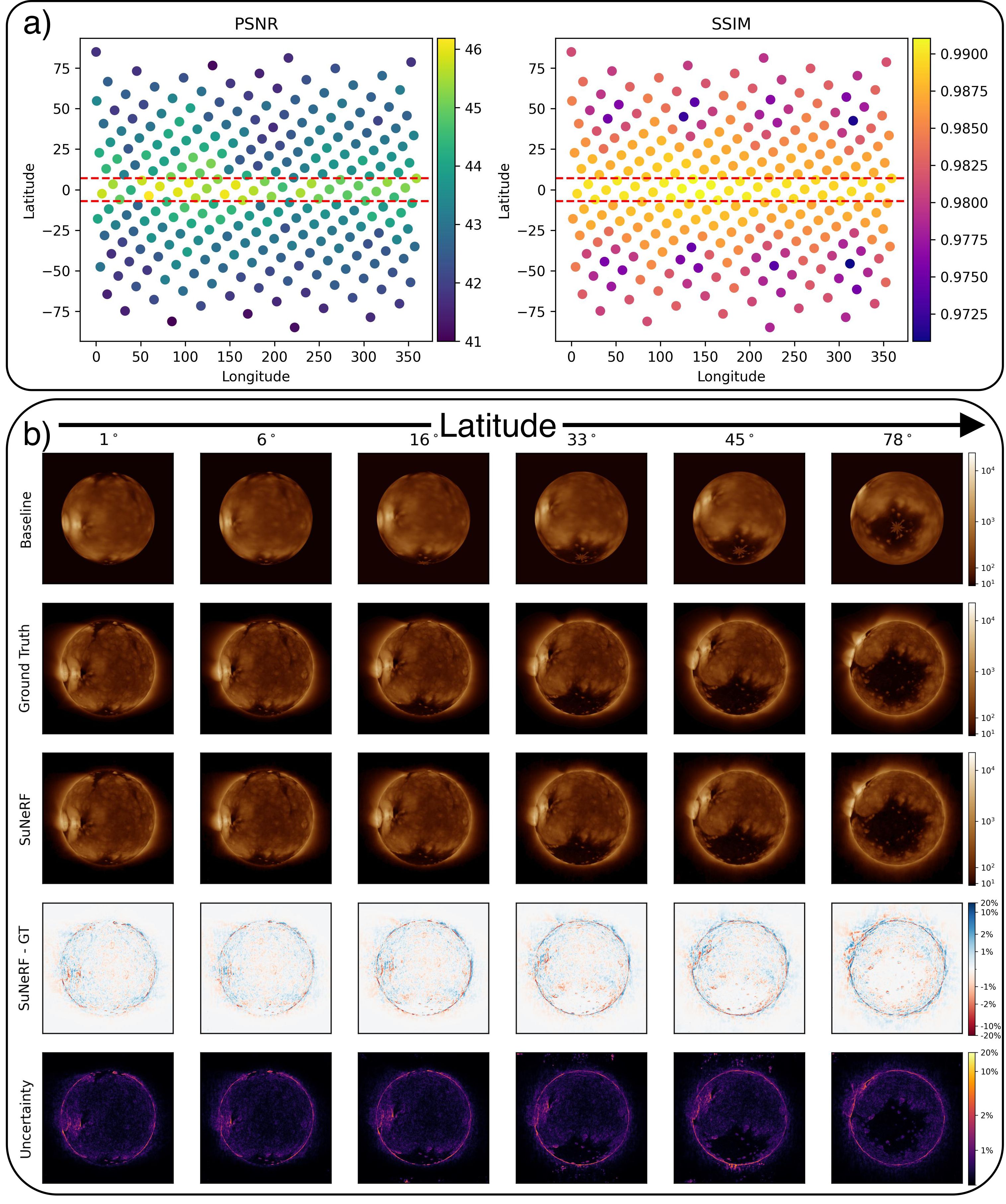}
    \caption{Evaluation of the SuNeRF using forward-modeled data of the EUV Sun at 193{\AA}. (a) PSNR and SSIM evaluated at 256 viewpoints, represented by points at the corresponding latitudes and longitudes. The color indicates the quality of the reconstruction, with larger values indicating a better agreement with the ground truth. Red dashed lines at latitudes of $\pm 7\degree$ mark the separation between the training and test viewpoints. (b) Qualitative comparison at different latitudes between the baseline method (spherical reprojection; first row), the simulation data (ground truth; second row), and the SuNeRF reconstruction (third row). Difference maps (fourth row) identify regions where our method deviates from the ground truth. Uncertainty estimates (fifth row) are in agreement with the errors. .
    } 
    \label{fig:results}
\end{figure}

\begin{table}[h!]
\caption{Metrics of performance for a synchronic map (baseline) and a SuNeRF applied to the test dataset (non-ecliptic viewpoints), including the peak signal-to-noise-ratio (PSNR), structural similarity index (SSIM), mean absolute relative errors (MAE), and mean relative errors (ME).}             
\label{table:evaluation}      
\centering                          
\begin{tabular}{| l || c c c c |}        
\hline
Method & PSNR (dB) & SSIM & MAE & ME  \\    
\hline                        
  Baseline - 193{\AA} & 23.6 & 0.68 & 3.4\% & -1.1\% \\
  \hline
  SuNeRF - 193{\AA} & 43.1 & 0.98 & 0.3\% & 0.05\% \\

\hline                                   
\end{tabular}
\end{table}

The endpoints of rays passing through the Sun are fixed to the surface (i.e., no emission from the interior), to account for the opaqueness of the solar surface (i.e., the transition to an optically thick medium). An overview of the SuNeRF model is depicted in Figure \ref{pipeline}.

\section{Results}
\label{results}
We trained our SuNeRF model for 120k iterations (approx. 30 epochs) using a batch size of 8096 rays ($\approx 19$ hours on a NVIDIA A100 GPU). We trained using backpropagation \citep[][]{Rumelhart1986LearningErrors} and the adaptive moment estimation (Adam) optimizer \citep[][]{Kingma2015Adam:Optimization} with a learning rate $lr = 5 \times 10^{-4}$, and we used mean squared error (MSE) as the loss function. Model uncertainties were estimated by fitting an ensemble of five SuNeRFs with different initializations and by computing the standard deviation of the outputs.

As a baseline, we generated a synchronic map from simulation images which we then reprojected to different viewpoints.  
Figure \ref{fig:results}(a) shows the peak-signal-to-noise-ration (PSNR) and structural-similarity \citep[SSIM;][]{wang2004ssim} for each viewpoint in the simulation. The distribution highlights that SuNeRFs provide throughout high-quality results, with a minimum SSIM of 0.97. Points close to the ecliptic show the smallest error, while the error gradually increases for higher latitudes, as expected from the train-test split. 
Figure \ref{fig:results}(b) compares our model to the baseline. At higher latitudes, the simple reprojections show artifacts and larger deviations from the ground truth, while the SuNeRF model renders almost identical images. Difference maps show that the primary errors occur close to the solar limb and off-limb region, which is also reflected in the uncertainty maps. We note that the reprojection method is incapable of handling the off-limb region.

Table \ref{table:evaluation} summarizes the quantitative evaluation over the full test set, with the SuNeRF model largely outperforming the baseline and showing no signs of over- or under-estimation.

\section{Conclusions}

In this work, we adapted NeRFs to generate a physically-consistent representation of the 3D Sun, with the inclusion of radiative transfer and geometric ray sampling that matches the physical reality of optically thin plasma. 
We trained a SuNeRF model using a 3D simulation of the EUV corona, with the training set limited to viewpoints that were captured from the ecliptic to reflect the limitations of current space-based observations. 
The remaining (non-ecliptic) viewpoints were used to evaluate the SuNeRF's ability to reconstruct the full 3D geometry of the simulation. 
A comparison to a baseline reprojection method shows that SuNeRFs provide new state-of-the-art results in 3D representations of the full Sun. 

Future work includes the expansion of SuNeRFs to other EUV wavelengths and to application to actual EUV observations to generate novel views beyond what existing satellites can achieve. 
Finally, we plan to expand SuNeRFs to output physical quantities such as plasma density and temperature (from which we then compute emission and absorption). 

\section*{Acknowledgements}

This work has been enabled by the Frontier Development Lab (FDL\footnote{Frontier Development Lab page: \url{https://frontierdevelopmentlab.org}.}). FDL is a collaboration between SETI Institute and Trillium Technologies Inc., in partnership with NASA, Google Cloud, Intel, NVIDIA and many other public and private partners. 
Any opinions, findings, and conclusions or recommendations expressed in this material are those of the author(s) and do not necessarily reflect the views of the National Aeronautics and Space Administration
The authors would like to thank the FDL organizers, the SETI institute, Trillium, the FDL partners and sponsors, and the reviewers for their constructive comments during the research sprint. 
The authors also want to thank Yarin Gal and Chedy Raissi for valuable discussions. Finally, the authors also thank Google and NVIDIA for providing access to computational resources without which this project would not have been possible. 

\newpage

\bibliography{neurips_2022}







\section*{Short statement}

In order to fully understand the physics that drives the dynamics of the Sun -- including its eruptive events and the resultant space weather in the solar system -- we must observe the Sun as it is: an evolving, three-dimensional star. To date, nearly all solar observations have been made along the Sun-Earth line, with the crucial notable exception of the twin STEREO satellites (pulling ahead and falling behind of Earth's orbit) that proved how valuable other viewpoints are. The next natural step in solar observations is to encircle the Sun with telescopes. The heliophysics community is calling for such a constellation. However, there is an inherent tension between the cost of adding more satellites to a constellation and the increased observational capability (science, forecasting) that results. The methods described herein alleviate that tension by enabling the addition of virtual observatories that can generate accurate synthetic images despite learning from a limited number of viewpoints (provided those viewpoints are sufficiently dispersed around the Sun). Once the model has been trained, there is no cost to adding virtual viewpoints at \textit{any} position around the Sun. This not only allows scientists to study dynamic phenomena from all angles, but these synthetic data can also be used to determine how Sunlight is distributed across the crucial parts of the spectrum that drive upper atmospheric chemistry or bombard the surface of planets with no atmosphere. This is unprecedented. To date, only Earth and Mars have measurements of both the solar input and the planet's atmospheric response; while most planets have measurements of the latter, the tools developed here are needed to provide the former. Ubiquitous coverage of the Sun means that better space weather forecasts can be produced for Earth and anywhere else. This capability is critical to develop and test now so that future crews en route to and at Mars can plan and act accordingly. 

\end{document}